# New Possible Mechanisms of Thunderstorm Clouds Physics

Manana Kachakhidze, Nino Kachakhidze-Murphy

Georgian Technical University, Faculty of Informatics and Management Systems

*Correspondence to:* kachakhidzem@gmail.com

**Abstract**

The work provides possible mechanisms of physics of thunderstorm clouds. We tried to consider the problems of the hail clouds in a completely different view. The article is based on the latest scientific achievement of modern theoretical physics, chemistry, physics of thunderstorms clouds and gives opportunities for new ways searching of impact on hail clouds.

**Introduction**

It is known that a strong earthquake preparation process can be accompanied by various geophysical anomalies, which expose themselves several months, weeks or days prior to earthquakes. Such as: Changing of intensity of electro-telluric current in focal area; Perturbations of geomagnetic field in forms of irregular pulsations or regular short-period pulsations; Perturbations of atmospheric electric field; Irregular changing of characteristic parameters of the lower ionosphere (plasma frequency, electron concentration, height of D layer, etc.); Irregular perturbations reaching the upper ionosphere, namely F2-layer, for 2-3 days before the earthquake; Increased intensity of electromagnetic emissions in upper ionosphere in several hours or tenths of minutes before earthquake; Lighting before earthquake; Infrared radiation; Total Electron Content (TEC) anomalies; Changing of weather parameters.

Physical mechanisms of the mentioned phenomena by us are explained on the base of the classical electrodynamics [16]. As it was expected, in the origination of the above mentioned anomalies the defining role electromagnetic radiation plays.

Exceptions are not the case of cloudiness intensification, which, of course, implies a good weather change with the "bad weather". This fact has led us to study more deeply the issue of "Bad weather", namely, the hail origination.

According to their electrical characteristics, clouds divide into thunderstorms and non-thunderstorms.

Typically, a thundercloud is a cumulonimbus cloud within which intra-cloud lightnings or cloud to ground disharges are observed. Physically, lightning is a short-term electrical discharge, the length of which is usually equal to several kilometers.

The presence of lightning - the main distinguishing feature of thunderstorm clouds. [8].

The thunderstorm cloud is a localized area of strongly marked convective and electrical activity. It can consist of one or more cells. The lifetime of the cell, from the moment of its nucleation to decay, is about 1 hour [24].

Usually the development of a thunderstorm cloud divides into three stages: nucleation, development (maturity) and decay. The stage of nucleation characterizes by the presence of sufficiently powerful ascending streams of warm humid air and the appearance of the first lightning. In the development stage, the electrical activity, the ascending streams and the moisture content of the cloud increase and in the decay stage there is observed attenuation of the ascending movements of the air, attenuation of electrical activity and precipitation [24].

**Discussion:**

In the physics of the formation of thunderstorm clouds, the main issues are:
- How the opposite charges appear inside of the microscale size clouds;

- How does occur a spatial macroscale (over a distance of several kilometers) separation of these charges, accompanied by the appearance of a strong electric field in the cloud;
- How are lightning discharges formed  [8];
- What is the mechanism of the creation of hail clouds;
- How does it happen the moisture accumulation, absorption and retention of large amounts of water in the cloud;
- How does it occur the origination of hailstones very fast, for about 30 minutes;
- What role do lightning and thunderstorms play?
- What is the decay mechanism of hail clouds?

But, before discussion of these major problems, we will briefly review the achievements of theoretical physics that have been used in different directions for 25 years.

In the last few years, a new section of electrodynamics - electrodynamics of materials with negative refraction, has been rapidly developing.

With negative refraction, light attracts the object to the source (emitter). Materials having negative refraction describe by a negative refractive coefficient n. By combining materials with negative and positive refraction, it is possible to organize the passage of a light ray through a set of material media by a certain law. That is, it can somehow turn, wriggle.

The phenomena associated with negative refraction observe not only in the sphere of electromagnetic radiation, but also in acoustics, in the propagation of various kinds of oscillations [46].

Refraction changes in materials with a negative refractive coefficient.

The decisive factor determining the dispersion of electromagnetic wave is the refraction coefficient. Also, a special property of such materials is the characteristic dispersion of waves propagating in such medium. Such medium are called metamaterials or "left-handed" medium [45].

To characterize the propagation of a wave and its dispersion, it is necessary to define the phase and group velocities of the wave. In the case of a plane harmonic wave, the phase velocity along the wave vector is the velocity of the motion of the equal phases surface.

Negativity of the phase velocity means, that when the wave propagates, the phases incursion take place in the direction from the receiver to the source, while the transfer of energy, in an obvious way, comes to pass from the source to the receiver.

Group velocity is a parameter characterizing the propagation of velocity of a "group of waves", that is, the propagation of a more or less well-localized quasimonochromatic wave (wave with a sufficient narrow spectrum). The group velocity usually interpret as the rate of displacement of the maximum of the amplitude enveloping of a quasi-monochromatic wave packet.

If the dispersion properties of the medium are such that the wave packet propagates in it without significant changes of enveloping shape, the group velocity can usually be interpreted as the velocity of the "energy transfer" of the wave or the velocity, with which signals can be transmitted by the wave packet carrying information in accordance with the principle of causality.

As follows from the theory of relativity, the group velocity is always positive and by numerical value is less or equal to the velocity of light in a vacuum. The group velocity by numerical value coincides with the phase velocity in one-dimensional medium without dispersion [45].

For materials with negative refraction:
1. The mass carried by the electromagnetic field in the matter from the emitter to the receiver expresses as:
$$m = \frac{E}{V_{ph}V_{gr}}$$

The relation $E = mc^2$ is a special case of the aforecited relation.

2. Inside the material with negative refraction, the light pressure is replaced by the light attraction, and the mass is transferred by light not from the emitter to the receiver, but on the contrary, from the receiver to the emitter.

3. The Abraham tensor is not relativistically covariant and in general it cannot be used to calculate forces, acting from the electromagnetic field on the matterial. These forces can be calculated on the base of the Minkowski tensor. [47].

Many theoretical and experimental articles are dedicated to the issue of metamaterials, where, basically, it is talking about using of this effect in technology [1, 22, 23, 44].

Later it turned out that the things with metamaterials properties exist in the nature as well. For example, the orphan stone or the colorful butterflies. It turns out that this is due to the fact that the structure of semiprecious stones and wings of butterflies is practical identical: they consist of periodically located elements, some formations that cause light to propagate and reflect on completely unusual directions, forming various color gamma. [30].

Because in the atmosphere rarely but still takes place a negative refraction [12, 20, 26] in our opinion, the atmosphere, in certain conditions, acquires metamaterials properties.

Let us recall for a more profound discussion of the issue that a crystal is a solid body in which the atoms are periodically arranged, causing the movement of electrons along completely unusual directions.There are so-called forbidden zones for energies, where electrons cannot exist in a given area but can be only in allowed areas.

This leads to the existence, for example, of semiconductor crystals, a conduction band, a valence band, a forbidden band, where the energy for electrons is not allowed. But if this is not an intrinsic semiconductor but with impurities and some additives, these impurities allow electrons or holes to be captured by these impurities and create unusual states in the forbidden zone, entire subzones. Electrons can be captured, holes can be released, that is to form a new phenomenon for moving electrons.

It is known that electromagnetic waves have a corpuscular-wave dualism, that is, they are both particles and waves. The same can be said about the electron. An electron is a particle, but it has the properties of an ordinary wave. Therefore, electromagnetic waves have the properties of particles, such particles are called quasiparticles, in particular photons [30].

It is known that along with the increasing of altitude, the icy crystals appear in the atmosphere. Ice crystals in the atmosphere are generated as a result of the condensation of water vapor on aerosol particles and the heterogeneous crystallization of microaccumulations of supercooled water in the inhomogeneities of their surface [26, 36].

If we suppose that in the atmosphere, in certain conditions (because of electromagnetic waves with certain frequency, temperature, environmental orography, etc.) ice inclusions create natural periodic structures, (as it is in the opal and wings of mottled butterflies) i.e. they are crystals, which have exactly the same properties for electromagnetic waves or photons as ordinary crystals for electrons, then electromagnetic wave, prenetating it, would begin to propagate in completely unusual directions according to some special laws and all this would be connected as if it happens with electrons in real crystals [30].In these cases, these ice crystals present photonic crystals as well.

It should be taken into consideration that in this periodic structure exist some elements that will seize the wave even if the wave cannot propagate there. This is completely analogous to what happens in the ordinary crystals. Verification of this are the opals and butterflies wings, which consist of a periodic structure created by nature. As it was mentioned above, this structure let the light to propagate, be reflected or propagate according to completely unusual laws.

It is known that white light, that we see, consists of a spectrum: red, orange, yellow, green, blue, indigo and violet. The isolated elements - the spectral areas of this light - can be absorbed and can be reflected by a photonic crystal. Of course, such crystals are also created artificially.

Besides the photon crystals, there are magnetic materials in nature, where spin waves propagate - these are waves of the magnetic moment, their quasiparticles are called magnons.

Magnon crystals are also created artificially. These are periodic structures in magnetic materials that allow magnons to propagate according to laws that correspond to propagation of electromagnetic waves in photonic crystals.

There are also phonon crystals, i.e. periodic structures that allow acoustic waves to propagate in the phonon crystals according to laws that correspond to the laws of propagation of electromagnetic waves in photonic crystals. Phonon crystals are special elements, so-called wave filters, frequency filters, which sort out a certain frequency.

Thus, photonic and crystals similar to them can compel light or another wave, acoustic or magnetic, to propagate and reflect according to laws that contradict the laws of ordinary optics, acoustics or magnetism. Photonic and crystals similar to them also have the properties of metamaterials. [30].

Because the certain conditions can be created in atmosphere when generation of photonic crystals are allowed, it is not excluded that, in certain circumstances, the periodic structures may be created, which have properties of acoustic that is phononic crystals. This can prove to be the fact that scientists consider acoustic methods as one of the methods of influence on clouds. [26].

It's interesting, that people have mastered ways of fighting with hail. It is noticed that a shrill sound does not allow hailstones to form. Even the Indians preserved their harvest in this way - continuously hammering into the big drums as the thundercloud approached.

Our ancestors used bells for the same purpose [17].

**Possible mechanisms of thunderstorm clouds formation and decay**

As it was mentioned above, the periodic structures with metamaterials properties exist also in nature. At that, their properties do not depend on the chemical composition but depend on the geometric parameters of the medium elements [34].

Theoretical and experimental studies proved that negative reflection of ligth takes place in the troposphere [12, 20, 26,40]

Negative refraction is a rare phenomenon in the atmosphere and is observed with a sharp drop of air temperature by height (for example, in case of a thunderstorm) [49].

In our opinion atmosphere acquires the characteristics of the environment with a negative refraction only when a certain spectrum of electromagnetic waves propagate in it. Scientists, in the study of the thunderstorm and hail clouds, emphasize the fact that the thunderstorm and hail clouds are formed in certain places of the earth and mainly during a certain period of the day. It is not excluded that this fact is a defining factor by the view of the creation of periodic structures in the clouds.

Because the atmosphere characterize by negative reflections and there are structures in nature which in case of negative refraction have properties of metamaterial, it is not excluded that part of the atmosphere in certain cases will acquire properties of metamaterials.

Studies have shown that phase and group velocities have different directions in systems with negative refractive index. In addition, the energy that the group velocity transfers directs from receiver to the source [1, 22, 23; 44, 46].

If we consider in this regard the ongoing processes in the atmosphere, which are connected with the origination of hail clouds, according to the above mentioned theory, it is possible that clouds-clusters will be formed in this part of the atmosphere. In such case, the energy will be transferred by the light from the cloud to the emitter, because of which the energy inside the cluster will reduce. Cluster, which volume at the stage of maturity of a thundercloud practically unchanges, due to the pressure difference with respect to the environment, starts to work as powerful air sucking pump [8] , that's why it will start pumping the materials (water drops and at this time existing materials in the atmosphere) from the outside system in itself.

The mentioned cluster obviously is a locked system and due to comprehensible conditions, it consists of horizontal layers of clouds, which separates by some border zones from each other.

Despite such separation, this system is consolidated by the following factors:
1. The electromagnetic waves of certain frequency fall in exactly with that very "permissible" angle on these clouds system and changes its structure in that way, that the cloud, for its part, acquires properties of metamaterials;

2. All the layers of clusters are periodic systems, in the formation of which except of electromagnetic waves and temperature, the orography of under cloud cluster, plays also an important role;

3. At this time the strong inversion of electric field of atmosphere takes place, by that time the atmospheric electric field potential gradient reaches $10^5$ -$10^6$ V/m  [14].

In case if all three above mentioned conditions are fulfilled, noted system is enough strong and as we have mentioned, it manages to pump a large amount of water in itself like "sucking pump".  In the process of development of such cloud, a large amount, up to $(2 \div 4) \cdot 10^5$ tons of water, rises from the earth surface layer of the atmosphere to a height of several kilometers during the time less than 1 hour [39].

It is important to note that clouds are colloidal systems. In addition, the cloud is a dielectric and  this system represents an active environment during of water mass pumping.

It is clear, that after some time when the system loses its stability, it starts to collapse.

In the case of clouds this can happen by changing of the orography under the cloud, the angle of the sun radiation, the temperature, the height of the cloud, etc. That is, this periodic structure cannot exist long enough since the parameters necessary for creating such system change rapidly towards to moving clouds.

Of course, this  briefly described  process is devided into stages of origination, maturity and decay  of the thunderstorm clouds.

Let's go back to the initial stage of clouds creation: as soon as the periodic structure formation begins, the global electric field's circuit derange, an inversion of the atmospheric electric field occurs and because of water molecules polarization, acumullation of the charges starts inside of created cluster, in different layers.

In this case, the voltage of the electric field of the atmosphere between the earth and the ionosphere is directed from the earth to the ionosphere but in the cloud itself, it will be directed to the opposite direction. In the case of "bad weather", the fact of strong inversion of the atmosphere electric filed is experimental proven.  At this time the gradient of atmospheric electric field changes significantly, it varies between $10^5$ - $10^6$ V/m  [48].

It should be mentioned that the picture of the atmospheric electric field during the "bad weather" reiterates a picture of the distribution of the atmospheric electric field of the "bad weather" caused by the large earthquake preparation process [16]. But, in case of an earthquake, the primary role plays perturbation of the telluric current, which is caused by an earthquake preparation process and causes inversion of the atmospheric electric field. At this time the filtered quantities of atmospheric electric field's potential  gradient vary between $(-2500)$ V/m and 3000 V/m limits  [15].

It is possible that in case of "bad weather", the telluric current is excited also, but this will not be caused by the processes taking place in the earth crust, which always accompany earthquake preparation process, but by perturbation of atmosphere electric field, which in its turn is caused by the solar radiation.

It should also be noted that during the strong earthquakes only the cloudiness does not increase but some weather parameters (wind, humidity and etc.) changings also take place [21].

As for the hail (thunderstorms) clouds, the possibilities of its development depends not only on a certain spectrum of solar radiation, but also on the orographic conditions where hail is formed.

Obviously, the strong inversion of the atmospheric electric field, which takes place also in the case of hail clouds, is one of the main conditions of charges distribution and separation from each other. In addition, as it was mentioned, the field in the cloud will be directed down from the top. Studies have shown that this field enhances the coagulation process of the droplets and thereby intensifies the charge separation process [8].

In our opinion, the causes of the atmosphere ionization and hail clouds origination, in general, should be considered in the background not only of the metamaterials properties of the clouds but also on the background of the possibility of existence of carbon nanotubes in the atmosphere as well.

In this regard it is interesting article, in which the problem of transferring of electromagnetic radiation in optically dense multiphase media with multiple scattering is considered. Solutions for clouds of various geometric shapes consisting of particles with different optical properties are contemplated. In the paper are considered issues related to the propagation of electromagnetic waves through optically dense aerosol formations.

It is seen that liquids based on water with carbon particles have optimal characteristics. Increasing the volumetric content of carbon particles (the maximum value corresponds to a dense package and equals to 0.7) one can practically achieve the same effect as in the case of using carbon powder [18].

Thus, as seen from this work, the role of carbon is quite significant in the atmosphere.

Of course, the immediate cause leading to the formation of any clouds, including thunderstorms, is the condensation of water vapor contained in the air, in a result of which droplets appear in the atmosphere. This process occurs in the presence of supersaturated vapor in the air and the so-called condensation nuclei [8].

But in our view, the origination of the hailstorm clouds in the atmosphere has a much rough reason than it was described in the existing works. That is why we will try to describe this process in more detail: In particular, as soon as the periodic structures start to form, the clusters with low energy zones will appear in the atmosphere cloud. At the beginning these clusters will have very small sizes and if the conditions are not set up to strengthen these periodic structures, it will obviously fall apart.

Of course, this is already a nucleus of the thunderstorm cloud and this can happen on any height from the earth's surface, including zero isotherm. Because of the weakness of the electric field, these clusters cannot propagate vertically yet and their shape will be basically flat.

In the cluster the low energy zones originate since this cluster has already acquire properties of metamaterials, because of it the energy is transferred from the cloud to the emitter.

The fall of energy will cause origination of ice sleet. Since the electric field of atmosphere mainly is not yet finally inversed at this time and the field's potential gradient is not high, it is not excluded that the process of separation of electric charge might be very weak.

It is clear, that the separation and movement of electric charges inside and around a thunderstorm cloud is a complex continuously changing process. Nevertheless, it is possible to imagine a generalized picture of the distribution of electric charges at the stage of cloud maturity [48, 51].

At the next stage, in the atmosphere emerged clusters will combine similar to each other periodic structures, some of which are composed of the smallest icy crystals. In the crystals, as noted earlier, there are so-called forbidden zones for energies where electrons cannot exist in a given areas, but can be only in permitted areas. Therefore, from these clusters, under the influence of the electromagnetic waves of certain frequencies electrons will be expelled and they will stay in the external space of given cluster. But since strong inversion of the electric field of atmosphere is already exists and in parallel convection takes place, clusters with positive charge will climb to the top in the cloud (at this time, the potential of the ionosphere is negative). By then, the expelled electrons will be dropped down under the influence of the inversing electric field (Earth's potential, in this case, is positive). That is why, usually, a positive dipole structure dominates, in which the positive charge is in the upper part of the cloud, and the negative charge is under it inside the cloud. [48, 51].

Of course, in parallel, during this time the natural ionization takes place everywhere in a result of constant effect of various types of radiation (cosmic, ultraviolet, ionizing) and atmospheric electricity [19]. Atmospheric ions, moving under the action of an electric field, form shielding layers on the boundaries of the cloud that mask the electrical structure of the cloud from an external observer [3, 43].

Thus, the charges of different signs will be redistributed in the opposite parts of the cloud and this distribution of the charges cannot be disturbed until such a strong inversion of the electric field of the atmosphere takes place. However, this process is anomalous and cannot last long. The moment will come when the system will weaken, i.e. the electric field potential gradient will decrease and in the lower part of the cloud, under the layers of electrons, will appear particles with positive charge.

As for clouds clustering, it is permissible that outwardly uniform cloud would be composed of separate layers with different charges, among which the atmospheric discharges take place. Strong clustering structures that are already forming as uniform system at the maturity stage of the thunderstorm clouds and have a constant volume in a certain period, as mentioned above, reveal properties of the "suction pump" and transfer materials that are outside of this cluster, into its interior. That's why in the process of development of such a cloud, from the earth surface layer of the atmosphere to a height of several kilometers, in a time less than 1 hour, a large amount of water, up to $(2 \div 4) \cdot 10^5$ tons, rises [39].

The thunderstorm cloud always has greater vertical extent and moisture content, than usual (non-thunderstorm) cumulonimbus cloud. This means that stronger upwelling streams of air exist in a thundercloud, throwing moisture at a high altitude.

The thunderstorm cloud, like an air pump, pushes humid air from the earth surface layer of the atmosphere to a much higher altitude than the usual cumulonimbus cloud. The average radius of the base of one thunderstorm cell is $R \approx 2$ km, in the middle latitudes the peak of a typical cell is located at altitudes $(8 \div 12)$ km [24].

In order to form an average thunderstorm cloud with a moisture content of $2 \cdot 10^5$ tons and with a center of gravity located at an altitude of 4 km, it is necessary that a moist air mass of $\sim 10^7$ tons be lifted from this earth surface layer to this height. At the same time, the energy expenditure for overcoming gravity should be $\sim 4 \cdot 10^{14}$ Joules. In the process of formation of such a cloud, during the condensation of steam into atmosphere, $\sim 5 \cdot 10^{14}$ J of energy (latent heat) should be released. It is the very energy which is spent on the formation of a thundercloud. [8].

The current flowing through a medium-sized cloud is $\sim 1$ A. The voltage between its vertex and the base is $(10^8 \div 10^9)$ V, and the electric power is $\sim (100 \div 1000)$ MW. [11, 35, 42].

The main distinguishing feature of a thundercloud from any non- thunderstorm cloud is the presence of lightning in it [8].

As soon as the value of $E$ in the cloud reaches the values (2 - 3) kV / cm, lightning appears in it. Measurements of the value of $E$ performed on cylinders showed that in thunderstorm clouds, as a rule, the values of $E$ are less than 2 kV/cm [25]. Measurements on the aircraft showed that the maximum values of $E$ did not exceed 3 kV/cm [9]. These maximum values were observed at the time of the appearance of the lightning discharge [8].

Apart from intra-cloud lighting there are also ascending and descending types of lightning (cloud-to-ground discharges). Ascending lightning appear at the last stage of the development of the cloud, when its water content is maximal. It's important that when the frequency of occurrence of lightning varies from 0 to 20 lightning / min, the water content of the cloud varies from 50 to 350 thousand tons [8].

As soon as the lightning flashes, the negative charge is extinguished and the hailstones begin to fall [6] The hail almost always falls exactly before the downpour or simultaneously with it and never after it. The average duration of hailstorms is from 5 to 20 minutes [14]. This already is an irreversible process and the hail clouds do not occur again after hail.

It is proven, that in any region of the globe, a maximum of thunderstorm activity is observed at about 4 pm local time (LT). At this time the most intense evaporation of moisture and updraft from the surface of the Earth take place [4].

Besides the electromagnetic waves with specified frequency, orography, cloud thickness and its location towards solar disk, play an important role in the formation of hail clouds.

Another problem that is related to the hail origination is the creation of hailstones itself and their keeping problem in the atmosphere for some time. As noted above, for the formation of hailstones mandatory requirement is the existence of condensation (crystallization) centers.

As soon as supersaturated vapor appears in the air rising from the earth surface layer of the atmosphere, the process of cloud formation begins (condensation of moisture). The role of condensation nuclei is performed by wettable (nonfat) aerosol particles [8].

The main supplier of uncharged aerosol particles in a thundercloud is air rising from the earth surface layer of the atmosphere. In turn, the main supplier of light ions in a thundercloud are lightnings. As light ions stick to aerosol particles, thanks to lightning, a large number of charged aerosol particles are formed in a thunderstorm cloud. These particles are the most active nuclei of vapor condensation [8].

The number of light ions formed in a thunderstorm cloud is mainly determined by the electrical activity of the cloud (by the number of lightning per unit time). This activity depends on the degree of air ionization in the earth surface layer of atmosphere, as well as on the degree of ionization of the air column [8].

It is known from observations that when a thundercloud is in the maturity stage, the average time of electrical activity of one thundercloud continues (20–30) min. Condensation of vapor occurs mainly during the electrical activity of the cloud. In a cloud with a base of ~ 13 km$^2$, the frequency of lightning discharges is about 30 bits in second [29].

The existence of the saturated vapor in the atmosphere, the origination of ions and the role of lightning in hail origination, described above, are a necessary conditions and represent the contributing factors of the hail formation in the atmosphere, but we have different opinion about the origination of cloud condensation nuclei and directly about related to the formation of hailstones. Therefore, in addition to the above mentioned, we also rely on the recent published works in scientific literature, for example:
- Hail usually takes place in locations where the strong discharge of lightning happens and is always associated with a thunderstorm. There is no hail without a thunderstorm [14].

- Fullerenes exist in nature everywhere there is carbon and high energies (volcanoes, lightning), are formed by gas burning in the home gas stove or in the flame of an ordinary lighter, their skeleton consists of carbon atoms, which has a high adsorption capacity [53].
- Fullerenes are an effective trap for free radicals [33, 53]
- Fullerenes are also found out in wildlife. [5, 32]. So, radiolarians, a subclass of unicellular animals of the sarcodic class, have skeletons of various geometrical, including fullerene forms [27].
- Fullerenes are formed spontaneously at high temperatures. The simplest molecule consists of 60 carbon atoms which are composed of hexagons and pentagons of carbon, joined together to form a completely spherical shape [2, 38].
- Fullerenes have many interesting physical properties. For example, the fullerene contained in shungite contributes to the ordering of water structures and the formation of fullerene-like hydrate clusters in it [33].

At temperatures below 0 ° C, fullerenes are transformed into a cubic lattice [31].

It should be emphasized that in recent works the authors directly point out to the presence of fullerenes in the atmosphere:
- In the summer of 2011, the results of air sampling over the Mediterranean Sea were published: fullerenes were found in all 43 air samples taken from Barcelona to Istanbul [37].
-It is interesting also the work in which the transformation of $C_{60}$, as nanocrystal ($nC_{60}$) aerosols, is evaluated over a range of simulated atmospheric conditions [7].

It is known that during condensation so-called condensation nuclei begin to form in the air. They are accumulations of water vapor molecules with reduced kinetic energy. If such nuclei are stable, they turn into droplets and crystals suspended in air or precipitated on surfaces. The formation of nuclei also requires the presence of condensation nuclei. Condensation is the only process of forming clouds of any shape [10].

It should also be taken into consideration, that when a lightning is discharged, as a result of receiving a huge amount of heat, the water along the channel of the lightning discharge or around it, rapidly evaporates; as soon as the lightning flashes cut off, it (water) starts to cool down very much. The evaporation process consumes the kinetic energy of this polarized water system itself.

In this process, strong and instantaneous evaporation ends with a strong and rapid solidification of water. The stronger the evaporation, the more intense the process of water solidification. For such process it is not necessary that the ambient temperature be below zero.

When lightning is discharged, various types of hailstones are formed, differing in size as well. The size of the hailstone depends on the power and intensity of the lightning. The more powerful and intense the lightning, the larger the hailstones are. Usually, hailstones precipitate quickly cease as soon as the flashes of lightning cut off [14].

Thus, according to the above stated, it is possible to form the following scheme: fullerenes appear in the already formed clouds where exists the oversaturated steam, on the background of high temperatures, during heavy lighting and fullerene-like hydrated clusters, part of which, according to the above-mentioned work, will be transformed into (nC60) aerosols, which in turn, represent the condensation centers. Because the fullerenes link up to free radicals pretty well [33], they may quickly join the charged water drops and because of falling of temperature after lightning, they also quickly produce hailstones.

It should be also noted that fullerene-like structures retains their properties for only a few hours [28, 41] which, in turn, promotes rapid disintegration of hail clouds as soon as the hail originates. This fact is observed by experiments as well.

It is also interesting the forms of hailstones.

One of the most common forms of hailstones are conical or pyramid form with sharp or slightly truncated tops and a rounded base.

The upper part of such hailstones is usually softer, opaque as if snowy; The middle part is translucent, consisting of concentric, alternating between transparent and opaque layers. The lower part is widest and transparent. Not less often a special shape is observed consisting of an inner snow core (sometimes, although less often, the central part consists of transparent ice), surrounded by one or several transparent shells. There are also pellet hailstones, with deepenings at the ends of the minor axis, with various protrusions, sometimes crystalline (ice pellet with large scalenohedron accrued on them [52]. Only hexagonal prisms have a shape close to the shape of the particles of crystals in the clouds [12].

If we assume that fullerene has a one of the crucial role in the generation of hailstones and the spatial forms of the fullerenes create different geometrical figures, it is easy to explain the forms of hailstones [33].

According to the results obtained in the article, it is possible to make the following conclusions:

1. Rarely, but a negative reflection still takes place in the atmosphere;
2. Negative refraction means that the atmosphere may have already the properties of metamaterials in this area;
3. The change of atmospheric properties is associated with the formation of certain periodic structures under the influence of definite frequency electromagnetic (including light) and acoustic waves in certain orographic conditions;
4. From some clusters, under the influence of the electromagnetic waves of certain frequencies will expulse electrons and they will stay in the external space of the certain cluster;
5. In this part of the atmosphere the global electric circuit interrupt and it happens atmospheric electric field inversion;
6. The charges of different signs will be redistributed in the opposite parts of the cloud and it will stay like this until a strong inversion of the electric field of the atmosphere takes place;
7. Due to the changes in the atmosphere, the hail cloud-cluster originates, in which, as in the body with the metamaterials properties, the phase and group velocities move in different directions;
8. As soon as the periodic structures start to form, in the cluster, the energy will be transferred by the light from the cloud to the emitter, because of which the energy will reduce inside the cluster and due to the pressure difference with respect to the environment, it starts to work as powerful air sucking pump;
9. One of the main condensation nuclei in the hail cloud are the fullerenes or fullerene-like structures that origin because of lightning;
10. The fullerenes or fullerene-like structures originate only in case of lightnings and thunderstorms;
11. $C_{60}$ fullerenes transform into nanocrystal ($nC_{60}$) aerosols;
12. During lightning the water rapidly overporates from the claster of the cloud and temperature quickly falls down in it;
13. Since the fullerenes are capable of catching free radicals pretty well, they join charged water droplets existent in the air which, due to low temperature around them, will be quickly transferred into hailstones;
14. Reducing of the negative charges disrupts the electric balance in the united cluster of clouds because of lightnings and the hail cloud becomes weak. This process also is promoting by the fact that the cloud location is changing in time towards solar disk, due of which periodic structures begin to decay. At the same time, the antigravity forces, which retentions the hailstones in the cloud, reduce and hail comes;
15. Hailing means that the cluster-hail cloud is collapsed. This process is irreversible.